\documentclass[twocolumn,showpacs,,prb]{revtex4}
\usepackage{psfrag}
\usepackage{graphicx}
\usepackage{amssymb}
\usepackage{amsmath,amsfonts,latexsym}

\newcommand{\be}{\begin{equation}}
\newcommand{\ee}{\end{equation}}
\newcommand{\bea}{\begin{eqnarray}}
\newcommand{\eea}{\end{eqnarray}}

\newcommand{\p}{\partial}

\newcommand{\lp}{\left(}
\newcommand{\rp}{\right)}
\renewcommand{\phi}{\varphi}

\renewcommand{\vec}[1]{{\mathbf #1}}
  
\begin{document}

\title{Pattern Formation in Exciton System near Quantum Degeneracy}
\author{L.\,S.~Levitov$^1$,  B.\,D.~Simons$^2$, and L.\,V.~Butov$^3$}
\address{$^1$Department of Physics,
Center for Materials Sciences \& Engineering,
Massachusetts Institute of Technology, 77 Massachusetts Ave,
Cambridge, MA 02139}
\address{$^2$Cavendish Laboratory, Madingley Road, Cambridge CB3 OHE, UK}
\address{$^3$Department of Physics, University of California San Diego, 
La Jolla, CA 92093-0319}

\begin{abstract}
We discuss models
of the modulational instability 
in a cold exciton system in coupled 
quantum wells.
One mechanism involves 
exciton formation in a photoexcited electron-hole system
in the presence of stimulated binding processes which build up near
exciton degeneracy. It is shown that such processes
may give rise to Turing instability leading to a spatially modulated state.
The structure and symmetry of resulting
patterns depend on dimensionality and symmetry.
In the spatially uniform
2d electron-hole system, the instability leads to a triangular lattice 
pattern while, at an electron-hole interface, a periodic 1d pattern 
develops.
Wavelength selection mechanism is analyzed,
revealing that the transition
is abrupt (type I) for the uniform 2d system, and continuous (type II)
for the electron-hole interface.
Another mechanism that could possibly drive the instability involves 
long-range attraction of the excitons. We illustrate
how such an interaction can result from plasmon wind, 
derive stability criterion, and discuss 
likelihood of such a scenario.
\end{abstract}


\maketitle

\section{Introduction}

Recent experiments observed striking spatial pattern formation 
in two-dimensional cold exciton systems
in photoexcited AlGaAs/GaAs 
quantum well (QW) structures~\cite{Butov02,Snoke02,Butov03,Snoke03,Rapaport}. 
The photoluminescence (PL) patterns span macroscopic scales in 
excess of $100\,\mu$m and include concentric rings~\cite{Butov02,
Snoke02,Butov03,Snoke03,Rapaport}
as well as `bright 
spots'~\cite{Butov02,Butov03}.
In addition to that,
the electron-hole system exhibits an abrupt transition at 
ca.~$2\,$K in which the outermost ring `fragments' into regularly 
spaced beads of high PL intensity~\cite{Butov02,Butov03}. While the gross
features of PL have been explained within 
the framework of a classical transport theory, 
attributing the internal rings to nonradiative exciton transport
and cooling~\cite{Butov02} and the outermost rings and 'bright spots'
to macroscopic charge separation~\cite{Butov03,Snoke03,Rapaport},
the origin of the instability remains unidentified. 

Spatially modulated exciton density 
is not to be expected in QW
system designed so that excitons interact repulsively
as electric dipoles~\cite{Butov02} 
and thus do not form droplets~\cite{Keldysh}. 
The macroscopic character of ring fragments,
of  $10-30\,\mu{\rm m}$ in size
and containing about $10^4$ excitons each, 
as well as the
observed spatial modulation of the exciton density
which appears abruptly at a certain temperature, 
calls 
for an explanation involving a symmetry-breaking 
instability of a homogeneous state to a patterned state. Such behavior is
reminiscent of the instability predicted by Alan Turing~\cite{Turing52} to 
occur in a reaction-diffusion system.
Such an instability 
of a reaction-diffusion system, first predicted in 1952 by Alan 
Turing~\cite{Turing52}, is characterized by an intrinsic wavelength 
resulting from an interplay of reaction and diffusion processes. 
The Turing instability is known to occur in certain chemical 
reactions~\cite{Castets90,Kepper91,Ouyang91}, as well as in
biological systems~\cite{Murray89}.

In this work we propose a mechanism, based on the kinetics of exciton 
formation from optically excited electrons and holes, that can lead to 
a Turing 
instability in the exciton system. Interestingly, these kinetic 
effects 
build up and 
become especially strong 
in the regime near \emph{exciton quantum 
degeneracy}, due to stimulated enhancement of the electron-hole binding
rate. 
The transition to a state with a spatially modulated 
exciton density reveals itself in the spatial PL pattern, and presents a
directly observable signature of degeneracy. 

Although, in itself, an 
observation of an instability does not constitute unambiguous evidence
for degeneracy, it may complement other manifestations 
discussed in the 
literature, such as changes in exciton 
recombination~\cite{Butov99} and scattering~\cite{Butov01} rates,
in the PL spectrum~\cite{Timofeev}, 
absorption~\cite{Johnsen01} and PL angular distribution~\cite{Keeling03}.
%
%
While linking the observed instability with degeneracy is 
premature, our main aim here is to present the Turing 
instability from 
a broader viewpoint,
as a novel and quite general 
effect of quantum kinetics that can help to identify the regime of 
Bose-Einstein condensation (BEC).


Besides the kinetic mechanism, we consider another 
mechanism based on the assumption that 
the interaction between excitons is of an attractive character 
at large distances. The origin of such an attractive contribution
may be in the force on excitons due to plasmons or phonons generated 
at exciton recombination. We describe a simple model
for the plasmon wind effect, analyze the stability criterion, and 
consider the possibility that the observed periodic patterns are 
related with such an attraction.

\section{Kinetic instability}
\subsection{Model}

Following Ref.~\cite{Butov03}, 
here we consider a transport theory~\cite{Butov03} formulated in 
terms of electron, hole, and exciton densities $n_{\rm e,h,x}$ obeying a 
system of coupled nonlinear diffusion equations:
%
\begin{eqnarray}
&& {\rm (e)}\quad \p_t n_{\rm e}=D_{\rm e}\nabla^2 n_{\rm e}-w n_{\rm e}
n_{\rm h}+J_{\rm e}\nonumber\\\label{eq:Diff}
&& {\rm (h)}\quad \p_t n_{\rm h}=D_{\rm h}\nabla^2 n_{\rm h}-w n_{\rm e}
n_{\rm h}+J_{\rm h}\\\nonumber
&& {\rm (x)}\quad \p_t n_{\rm x}=D_{\rm x}\nabla^2 n_{\rm x}+w n_{\rm e} 
n_{\rm h}-\gamma n_{\rm x}.
\end{eqnarray}
The nonlinear couplings account for exciton formation from free 
electron and hole 
binding.
Here, no attempt has been made to describe the
detailed and complicated density and temperature dependence of the physical 
parameters entering the model, nor to account for 
nonequilibrium
exciton energy distribution and
cooling due to phonon emission~\cite{Ivanov,Butov01}.
Instead, we adopt a more phenomenological approach and assume that, 
as a result of cooling, the system can be described by an effective 
temperature, which leaves the densities $n_{\rm e,h,x}$ as the only important 
hydrodynamical variables. The sources $J_{\rm e}$ and $J_{\rm h}$ in 
Eq.~(\ref{eq:Diff}) describe the carrier photo-production,
as well as the leakage current in the QW 
structure. 

In general, one can expect the electron-hole binding rate $w$ 
and, to a lesser extent, the exciton recombination rate $\gamma$ 
to depend sensitively on the local exciton density $n_{\rm x}$. 
Of the several mechanisms that could lead to such a dependence 
at low temperatures close to exciton degeneracy, perhaps the most 
important in the present context involve stimulated electron-hole 
binding processes mediated by phonons. These 
processes enhance the binding rate by a factor $f=1+N_E^{\rm eq}$, where 
$N_E^{\rm eq}$ denotes the occupation of exciton states. 
In thermal equilibrium, and at low temperatures, 
one can ignore the reverse processes of exciton dissociation: 
The binding energy, carried away by phonons, is much larger
than $k_{\rm B}T$.

For a degenerate exciton gas with $N_{E=0}>1$, the dominant process 
involves scattering into the ground state and the stimulated enhancement 
factor is expressed as
\begin{equation}\label{eq:f-factor}
f=e^u \,,\quad 
u\equiv \frac{n_{\rm x}}{n_0(T)}\,,\quad n_0(T)=\frac{2 g m_{\rm x}
k_{\rm B}T}{\pi\hbar^2}.
\end{equation}
Here $m_{\rm x}\simeq 0.21m_0$ represents the exciton mass, and $g$ denotes 
the degeneracy (in the indirect exciton system, the exchange interaction is 
extremely weak, and $g=4$). Equivalently, when reparameterized through its 
dependence on temperature, $u\equiv T_0/T$ where $T_0=\left(\pi\hbar^2/2
gm_{\rm x}k_{\rm B}\right) n_{\rm x}$ is the degeneracy temperature. At  
$T\sim T_0$ (equivalently $n_{\rm x}\sim n_0$), there is a crossover from 
classical to quantum Bose-Einstein statistics, and the stimulated enhancement 
factor $f$ increases sharply. 

Qualitatively, the stimulated transition mechanism for hydrodynamic
instability can be understood as follows:
A local fluctuation in the exciton density leads to an increase in the 
stimulated electron-hole binding rate. The associated depletion of the 
local carrier concentration causes neighboring carriers to stream 
towards the point of fluctuation presenting a mechanism of positive
feedback.
The wavelength, determined by the most unstable harmonic of the density,
characterizes the length scale of spatial modulation in the nonuniform state.

Before turning to the analysis of instability, it is useful to discuss 
intrinsic constraints on the dynamics (\ref{eq:Diff}) due to electric 
charge and particle number conservation. These are obtained by considering 
the linear combinations ${\rm (e)-(h)}$, ${\rm (e)+(h)+2(x)}$ of the 
transport 
equations (\ref{eq:Diff}). In both cases, the nonlinear term drops out and 
one obtains linear equations
\begin{eqnarray}\label{eq:e-h}
&& \hat L_{\rm e} n_{\rm e} - \hat L_{\rm h} n_{\rm h}
=J_{\rm e}-J_{\rm h}
\\ \label{eq:e+h+2x}
&&
\hat L_{\rm e} n_{\rm e} 
+ \hat L_{\rm h} n_{\rm h}
+ 2\hat L_{\rm x} n_{\rm x} +2\gamma n_{\rm x} = J_{\rm e}+J_{\rm h}
\end{eqnarray}
with 
$\hat L_{\rm e(h,x)}=\p_t-D_{\rm e(h,x)}\nabla^2$. Note that, since the 
origin of the relations (\ref{eq:e-h},\ref{eq:e+h+2x}) 
is routed in conservation laws, they are robust and 
insensitive to the exact form of the electron-hole binding term.

\subsection{The two-dimensional problem}

\begin{figure}[hbt]
\psfrag{(k/l)**2}{$({\bf k}/k_\ast)^2$}
\psfrag{u>uc}{$u>u_c$}
\psfrag{u<uc}{$u<u_c$}
\psfrag{u=uc}{$u=u_c$}
\psfrag{k1}{${\bf k}^{(1)}$}
\psfrag{k2}{${\bf k}^{(2)}$}
\psfrag{k3}{${\bf k}^{(3)}$}
\centerline{
\includegraphics[width=0.75\linewidth,angle=0]{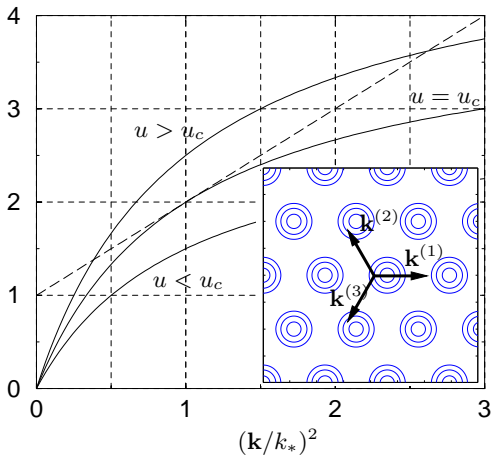}}
\vspace{-0.1in}
\caption{Graphical solution of Eq.~(\ref{eq:f(k)=0}) that selects the most 
unstable wavelength. {\it Inset:} The $3$-fold symmetric star of wavevectors 
describing the modulation near the instability threshold and the 
corresponding triangular pattern of exciton density variation.}
\label{fig:2d}
\end{figure}

The simplest to analyze is instability of a steady state
with spatially uniform carrier sources and density distribution,
which can be achieved in the presence of spatially extended photoexcitation,
described by constant $J_{\rm e}(r)=J_{\rm h}(r)\equiv J$
in Eqs.(\ref{eq:Diff}). 
In a uniform steady state, 
ignoring the dependence of the radiative 
recombination rate $\gamma$ on exciton 
density, we have 
\begin{equation}\label{eq:uniformstate}
\bar n_{\rm x}=J/\gamma \,,\quad \bar n_{\rm e,h}=\lp J/w(\bar n_{\rm x})
\rp^{1/2}.
\end{equation}
The stability of the system
can be assessed by linearizing Eqs.~(\ref{eq:Diff}) about the uniform 
solution (\ref{eq:uniformstate}) with a harmonic modulation 
$\delta n_{\rm e,h,x} \propto e^{\lambda t}e^{i\vec k\cdot\vec r}$. 
Using 
(\ref{eq:e-h},\ref{eq:e+h+2x}) and writing $\hat L_{\rm e}\delta 
n_{\rm e}=\hat L_{\rm h}\delta n_{\rm h}=- \hat L_{\rm x}\delta n_{\rm x}
-\gamma \delta n_{\rm x}$,
one can express $\delta n_{\rm h,x}$ in terms of $\delta n_{\rm e}$ and obtain
\begin{equation}
L_{\rm e}(\lambda,\vec k)+\gamma\frac{\bar{n}_{\rm x}}{\bar{n}_{\rm e}}
\lp 1+\frac{L_{\rm e}(\lambda,\vec k)}{L_{\rm h}(\lambda,\vec k)}\rp =
\frac{\gamma u L_{\rm e}(\lambda,\vec k)}{L_{\rm x}(\lambda,
\vec k)+\gamma}
\label{eq:2Dlinearized}
\end{equation}
where 
$u=d\ln w/d\ln \bar{n}_{\rm x}$ is
evaluated 
at the steady state (\ref{eq:uniformstate}), and $L_{\rm e(h,x)}(\lambda,
\vec k)=\lambda+D_{\rm e(h,x)}\vec k^2$. Solving Eq.~(\ref{eq:2Dlinearized}),
one obtains the growth rate dispersion $\lambda(\vec k)$.
The stability criterion ${\rm Re}\,\lambda <0$ is first violated at the 
parameter values such that the sign of $\lambda(\vec k)$ reverses  
for some $\vec k$. The condition $\lambda(\vec k)=0$ gives 
\begin{equation}\label{eq:f(k)=0}
({\bf k}\ell_{\rm x})^2+r=u\frac{({\bf k}\ell_{\rm x})^2}{1+
({\bf k}\ell_{\rm x})^2},
\end{equation}
where $\ell_{\rm x}=\sqrt{D_{\rm x}/\gamma}$ denotes the exciton diffusion
length and $r=D_{\rm x}(D^{-1}_{\rm e}+D^{-1}_{\rm h})\bar{n}_{\rm x}/
\bar{n}_{\rm e}$. Eq.~(\ref{eq:f(k)=0}) has solutions if 
%
\begin{equation}\label{eq:dlnw/dlnn}
u\equiv \frac{d\ln w}{d\ln \bar{n}_{\rm x}} \ge u_c=\lp 1+r^{1/2}\rp^2
\end{equation}
as illustrated in Fig.~\ref{fig:2d}. At $u=u_c$, we obtain the most unstable 
wavenumber $k_\ast = r^{1/4}/\ell_{\rm x}$
selected by competition of the stimulated binding and diffusion 
processes. 

The binding rate $w(n_x)$ builds up near degeneracy due to the 
growth of stimulated processes, leading to instability
at temperatures approaching $T_{\rm BEC}$.
(For a weakly interacting Bose gas, the transport coefficients, and thus 
the constant $r$, are 
practically insensitive to the degree of exciton degeneracy 
at $T>T_{\rm BEC}$.) 

To what extent are these results insensitive to the origin of the nonlinearity
in the binding rate? If enhanced by intraband Auger processes, 
which transfer the binding energy released in exciton formation 
to other excitons, one expects the 
binding rate $w$ to scale linearly with local exciton density, viz. 
$w(n_{\rm x})=w_0(1+n_{\rm x}/\tilde n_0)$, where $\tilde n_0$ denotes some constant 
involving a ratio of the two-body and three-body cross-section of the electron 
and hole in the presence of excitons. Crucially, in this case, the left hand 
side of Eq.~(\ref{eq:dlnw/dlnn}) is bounded by unity, while the right 
hand side is in excess. Therefore, at least over the parameter 
range considered here, one can infer that a simple linear scaling of the 
binding rate with density 
does not lead to 
instability. Indeed, the instability may be used 
to discriminate against certain mechanisms in the kinetics of exciton 
formation.

Turning to the discussion of the spatial pattern resulting from the 
instability, we note that the wavevector selection determines its 
modulus, but not direction. At threshold $u=u_c$ all modes 
with $|\vec k|=k_\ast$ become unstable simultaneously. The resulting $2$d 
density distribution can be found by considering the effect of mixing  
different harmonics due to higher order terms in  
(\ref{eq:Diff}) expanded in $\delta n_{\rm e,x}$ about the uniform state. 
Since these equations contain quadratic terms, the favored combination of 
harmonics is a $3$-fold symmetric star 
\be
\vec k^{(j)}=
k_\ast (\cos(\textstyle{\frac{2\pi}3} j+\theta),
\sin(\textstyle{\frac{2\pi}3} j+\theta)),
\ee
$j=1,2,3$,
with the phase parameter $\theta$ describing the degeneracy with respect to 
$2$d rotations. This leads to a density distribution $\delta n\propto
\sum_je^{\pm \vec k_j\cdot\vec r}$ with maxima arranged in a triangular 
lattice. 

On symmetry grounds, since the triangular lattice pattern
is stabilized by quadratic terms, the mean field analysis predicts that
the transition to the modulated state in this case is abrupt, of a type I kind.
Indeed, the triangular lattice geometry arises in various 
$2$d pattern selection problems, from B\'enard convection cells~\cite{Benard} 
to the mixed state of type II superconductors~\cite{typeII}.

\subsection{Instability of a 1D electron-hole interface}

\begin{figure}[hbt]
\psfrag{x/l}{{$x/\ell$}}
\psfrag{U}{$U_{\rm eff}$}
\psfrag{Y}{$\psi\,(\times 10)$}
\centerline{
\includegraphics[width=0.7\linewidth,angle=0]{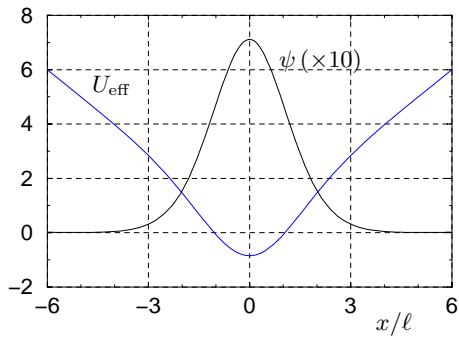}
}
\vspace{-0.1in}
\caption{Numerical solution of the (normalized) bound state wavefunction 
$\psi(x)$ 
representing fluctuation in the electron density, together with the 
effective potential $U_{\rm eff}=\bar{g}_{\rm e}+\bar{g}_{\rm h}-a_0
\,\bar{g}_{\rm e}\bar{g}_{\rm h}$ 
of the Schr\"odinger-like 
equation~(\protect\ref{eq:se}).\vspace{-0.1in}}
\label{fig:wav_pot}
\end{figure}

The application of these ideas to the $1$d modulation seen in 
exciton rings~\cite{Butov02,Butov03} was explored recently
in Ref.~\cite{turing_condmat}. 
The analysis starts with first determining the profile of the 
uniform distribution. The rings mark the interface 
between regions populated by electrons and holes at which they bind 
to form excitons. The steady state is maintained by a constant flux of 
carriers arriving at the interface. The parameter regime which 
is both relevant and simple to analyze is that of long exciton 
lifetime $\gamma^{-1}$ where the diffusion length $\ell_{\rm x}$
exceeds the range of the 
electron and hole profile overlap. In this case, approximating the source of 
excitons by a straight line
$c\delta(x)$, where $c$ is the total carrier flux and $x$ 
is the coordinate normal to the interface, the exciton density profile is 
given by $(c\ell_{\rm x}/2D_{\rm x})e^{-|x|/\ell_{\rm x}}$.
Accordingly, one can seek the electron and hole profile treating 
$w(n_{\rm x})$ as constant and restoring its dependence on $n_{\rm x}$ 
later when turning to the instability. The profiles can be inferred from 
two coupled nonlinear diffusion equations 
\begin{eqnarray}
\p_t n_{\rm e(h)} - D_{\rm e(h)}\partial_x^2 n_{\rm e(h)} = - w n_{\rm e}n_{\rm h},
\label{eq:Diff_eh}
\end{eqnarray}
with the 
boundary condition:
$D_{{}^{\rm e}_{\rm h}}\partial_x n_{{}^{\rm e}_{\rm h}}
|_{\pm\infty}=\pm c\,\theta(\pm x)$.
%
From Eq.~(\ref{eq:e-h}) one obtains $D_{\rm e}n_{\rm e}-D_{\rm h} n_{\rm h}=cx$
which allows the elimination of $n_{\rm h}$. 
Applying the rescaling $n_{\rm e(h)}=c\ell\, g_{\rm e(h)}/D_{\rm e(h)}$, 
where $\ell=(D_{\rm e}D_{\rm h}/wc)^{1/3}$, one obtains
%
\begin{equation}
\partial_{\tilde x}^2\ g_{\rm e}=g_{\rm e}(g_{\rm e}-\tilde x)
,\quad \tilde x\equiv x/\ell.
\label{eq:Diff_ge}
\end{equation}
From the rescaling one can infer that the electron and hole profiles 
overlap in a range of width $\ell\sim c^{-1/3}$ while 
\begin{eqnarray*}
g_{\rm e}(|x|\gg \ell) = \tilde x\,\theta(\tilde x)+{\cal O}\lp 
|\tilde x|^{-1/4}e^{-2|\tilde x|^{3/2}/3}\rp.
\end{eqnarray*}
The solution of this problem, obtained numerically, is 
displayed in Fig.~\ref{fig:wav_pot}.

Although Eqs.~(\ref{eq:Diff_eh}) are nonlinear, their diffusive character 
does not straightforwardly admit a spatial instability: A fluctuation in 
the position of the interface initiates an increased 
electron-hole flux which, in time, restores the uniform distribution. 
However, if one restores the dependence of the binding rate on exciton 
density, the same mechanism of positive feedback which characterized the 
instability in the uniform system becomes active.

To explore the instability, one may again expand linearly in
fluctuations around the spatially uniform solution, $g_{\rm e}(x,y)=
\bar{g}_{\rm e}(x)+\delta g_{\rm e}(x)\,e^{iky}$ (similarly $g_{\rm h}$ 
and $g_{\rm x}$), where $y$ is the coordinate along the interface
and $\bar{g}_{\rm e}(x)$ denotes the uniform profile obtained from 
Eq.~(\ref{eq:Diff_ge}). With $\ell_{\rm x}\gg \ell$, the exciton density 
remains roughly uniform over the electron-hole interface. Denoting this 
value by $\bar{n}_{\rm x}(0)$, in the vicinity of the interface, one may
again develop the linear expansion $w[n_{\rm x}]\simeq w[\bar{n}_{\rm x}(0)]
\,(1+u\delta n_{\rm x}/\bar{n}_{\rm x}(0))$ where, as before, $u=d\ln w/
d\ln n_{\rm x}$. Noting that Eq.~(\ref{eq:e-h}) enforces the steady state 
condition $\delta g_{\rm e}=\delta g_{\rm h}$, a linearization of 
Eqs.(\ref{eq:Diff})
obtains the Schr\"odinger-like equation
\begin{eqnarray}
\left[-\partial_{\tilde{x}}^2+(\ell k)^2+\bar{g}_{\rm e}+\bar{g}_{\rm h}\right]
\delta g_{\rm e}+\frac{u}{\bar{g}_{\rm x}(0)}\bar{g}_{\rm e}\bar{g}_{\rm h}
\,\delta g_{\rm x}=0,
\label{eq:se}
\end{eqnarray}
where 
the particle density Fourier components
$\delta g_{\rm x(e,h)}(q)=\int e^{-iq x}\delta g_{\rm x(e,h)}(x)dx$ obey the relation
\begin{eqnarray}
\delta g_{\rm x}(q)=-\delta g_{\rm e}(q)\left(1-\frac{1}
{Q^2+(q\ell_{\rm x})^2}\right), 
\label{eq:xe}
\end{eqnarray}
with $Q^2\equiv (k\ell_{\rm x})^2+1$. 
Now, since the product 
$\bar{g}_{\rm e}\bar{g}_{\rm h}$ is strongly peaked around the interface, 
the typical contribution from the last term in (\ref{eq:se}) arises from 
Fourier elements $q\ell\sim 1$. Then, with $\ell_{\rm x}\gg \ell$, the second 
contribution to $\delta g_{\rm x}(q)$ can be treated as a small perturbation 
on the first and, to leading order, neglected, i.e. $\delta g_{\rm x}\simeq 
-\delta g_{\rm e}$. In this approximation, the most unstable mode occurs at 
$k=0$. Qualitatively, an increase in $u$ will trigger an instability of the 
$k=0$ mode at a critical value $u_c$ when the linear equation first admits 
a non-zero solution for $\delta g_{\rm e}$. At the critical point, the 
corresponding fluctuation in the electron density then acquires the profile 
of the (normalized) zero energy eigenstate $\psi(x)$. Numerically one finds 
that the critical point for the instability occurs when $u_c/
\bar{g}_{\rm x}(0)=(2\ell/\ell_{\rm x})u_c \equiv a_0\simeq 6.516$, 
while the corresponding 
solution $\psi(x)$ is shown in Fig.~\ref{fig:wav_pot}.

While the approximation above identifies an instability, a perturbative 
analysis of the $k$-dependent corrections implied by (\ref{eq:xe}) reveals 
that the most unstable mode is spatially modulated. To the leading order of 
perturbation theory, an estimate of the shift of $u_c$ obtains 
\begin{eqnarray*}
\frac{\delta u_c}{u_c}=\frac{\ell}{a_0 a_1 }\left(
(k\ell)^2+\frac{a_0 a_2(Q)}{\ell_{\rm x}^2Q}\right)
\end{eqnarray*}
where $a_1=\int_{-\infty}^\infty dx\, \bar{g}_{\rm e}(x) \bar{g}_{\rm h}(x) 
\psi^2(x)\simeq 0.254\ell$, and 
\begin{eqnarray*}
a_2(Q)=\frac{1}{2}\int_{-\infty}^\infty dx dx'\, \bar{g}_{\rm e}(x) 
\bar{g}_{\rm h}(x) \psi(x) e^{-Q|x-x'|/\ell_{\rm x}} \psi(x').
\end{eqnarray*}
With $\ell_{\rm x}\gg \ell$, it will follow that $Q\ell \ll \ell_{\rm x}$, 
and the latter takes the constant value $a_2\simeq 0.461\ell^2$ independent 
of $k$.
%
%
Finally, minimizing $\delta u_c$ with respect to $k$, one finds that the 
instability occurs with a wavevector
\begin{equation}
k_c\ell_{\rm x}
\simeq \left(a_0 a_2\ell_{\rm x}/2\ell^3\right)^{1/3}
\label{eq:k_c}
\end{equation}
implying a shift of $u_c$ by 
%
%
$\delta u_c/u_c\sim (\ell/\ell_{\rm x})^{4/3}$. As a result, one can infer 
that the spatial modulation 
wavelength $\lambda_c\sim \ell^{1/3}\ell_{\rm x}^{2/3}$
is typically larger than the electron-hole overlap $\ell$, but smaller than  
$\ell_{\rm x}$.
Finally, 
an expansion of the nonlinear equations to higher order in fluctuations 
shows that, below the transition (i.e. for $u>u_c$), the amplitude of the 
Fourier harmonic $k_c$ grows as $(u-u_c)^{1/2}$. 

Once the instability is strongly developed (or when $\ell_{\rm x}\lesssim 
\ell$) the above linear stability analysis fails to produce quantitatively 
accurate results. Instead, one can perform numerical search for a steady 
state of the full nonlinear problem. Such an analysis, reported in 
Ref.~\cite{turing_condmat}, indeed demonstrates a modulational instability
developing in a soft way, with a square root dependence of the
modulation amplitude on the deviation from critical point
(see Fig.3 in Ref.~\cite{turing_condmat}).
The characteristic instability wavelength obtained numerically is consistent
with that predicted by 
Eq.~(\ref{eq:k_c}). 

\section{Long-range attraction mechanism}

Here we discuss the hydrodynamical forces between excitons that arise
due to polarization charge relaxation triggered by exciton recombination.
For the indirect excitons in $2$d coupled quantum wells 
we find a long-range attraction
that falls off very slowly as a function of distance. The strength
of this interaction is controlled by the specifics of 
the model, such as carrier mobility in the wells and adjacent doped regions.
Even if week, 
this interaction may overcome the repulsive
electric dipole interaction at a certain length scale, thereby making the
uniform state unstable with respect to density modulation formation.

The idea of the `plasmon wind' effect is as follows. Each exciton polarizes 
charges in the doped regions above and below the quantum wells. 
In the upper and lower doped region,
the polarization charges are of opposite signs,
forming a dipole that partially cancels the field of exciton dipole.
While this partial screening effect is small due to a large distance
to the doped region, and can be safely ignored in
the static picture, it acquires another dimension
and leads to a new effect if one takes into account that
excitons recombine at a constant rate. 

When an exciton recombines, its dipole disappears  
and the polarization charge, suddenly left to its own, starts spreading
in $2$d. The spreading is controlled by 
charge continuity and electro/hydrodynamics of the doped region. Thus 
an exciton cloud in which excitons disappear at a constant rate will 
represent a steady source of $2$d plasmons spreading radially away from 
it. The polarization charge in this situation is described by a $2$d 
Poisson equation with a source proportional to exciton density and,
therefore, the force on other excitons far away will be of a long-range 
$1/r$ form. The sign of this force is attraction, since the 
polarization charge dipole is opposite to that of exciton dipole. 

The significance of this force is that, while being weak compared to 
the direct 
exciton repulsion, it falls off slowly and comes to dominate the physics 
at sufficiently large distances. Being attractive, it favors exciton
clumping on large length scales. 
The situation can be understood by analogy with gravitation forces which are 
very weak, but long range, and thus are irrelevant on small length scales,
while controlling the structures forming on 
the astronomical length scales.

To estimate the effect of dynamical screening 
and plasmon wind on the interaction
between indirect excitons, we consider a simplified model. 
The excitons will be treated as electric dipoles positioned
in the $xy$ plane and oriented along the $z$ axis. 
The doped regions will be described as $2$d conducting planes
with conductivities $\sigma_{1,2}$, at a distance $a_{1,2}$
from the $xy$ plane and parallel to it.

The dynamical screening problem for indirect
excitons, described by dipoles $d$ with density $n_{\rm x}(\vec r,t)$,
is formulated in terms of the charge and current densities in the 
doped regions:
\be\label{eq:rho}
\p_t\rho_{1,2}+\nabla \vec j_{1,2}=s_{1,2}
\,,\quad
\vec j_{1,2}=-\sigma_{1,2}\nabla \phi_{1,2}
\ee
where $s_1=-\lambda(\rho_1-\rho_2)/2$, $s_2=-s_1$ describes leakage
through the structure.
The potentials $\phi_{1,2}$ are related to the charge density
in the two planes as
\be\label{eq:phi}
\phi_i=\sum_j U_{ij}(k)\rho_j+\phi_i^{(0)}
\ee
with $\phi_{1,2}^{(0)}$  
the exciton dipole potential in the planes
$z=a_1,-a_2$. Using
the Coulomb interaction $2$d Fourier components
\be
\frac{2\pi}{k}e^{-ka}=\int \frac{e^{-i\vec k\vec r}}{(\vec r^2+a^2)^{1/2}}d^2r
\ee
we obtain
\be\label{eq:Uij}
U_{i=j}(k)=2\pi/k
\,,\quad
U_{i\ne j}(k)=(2\pi/k)e^{-k(a_1+a_2)}
\ee
Similarly, $\phi_{1}^{(0)}=2\pi d e^{-ka_1}$,
$\phi_{2}^{(0)}=-2\pi d e^{-ka_2}$ 
is found for the potential in the planes $z=a_1,-a_2$
of a dipole $d$ situated at the origin.

To determine the force on a remote exciton, 
$\vec f=-\nabla U_{\rm eff}$, $U_{\rm eff}=-d\,E_z$,
we compute the electric field $z$ component
$E_z(k)=-2\pi(e^{-ka_1}\rho_1-e^{-ka_2}\rho_2)$
using $\rho_{1,2}$ obtained from 
Eqs.~(\ref{eq:rho}),(\ref{eq:phi}). In the simplest symmetric situation, when 
$\sigma_{1,2}=\sigma$ and $a_{1,2}=a$, we obtain
\be\label{eq:Ueff-general}
U_{\rm eff}(k,\omega)
=2\pi d^2  \lp \frac{2k^3 U_{12}}{2k^2\tilde U_{11}+(\lambda-i\omega)/\sigma}
+k \rp n(k,\omega)
\ee
where $\tilde U_{11}=U_{11}-U_{12}$ and $n(k,\omega)$ is exciton density. The second term corresponds 
to the direct exciton dipole interaction, while the first term describes 
the effect of polarization charge.

In order to extract the static interaction, as well as the part that depends
on the exciton recombination, we expand Eq.(\ref{eq:Ueff-general}) to the first order
in $\omega$ and obtain
\be\label{eq:Ueff(k,w)}
U_{\rm eff}(k,\omega)= (D_1(k)+i\omega D_2(k)) n_{\rm x}(k,\omega) 
\ee
with
\bea
&& D_1(k)= 2\pi d^2  \lp \frac{2k^3 U_{12}}{2k^2(U_{11}-U_{12})+\lambda/\sigma}
+k \rp 
\\
&& D_2(k)= 2\pi d^2  
\frac{2k^3 U_{12}}{\sigma(2k^2(U_{11}-U_{12})+\lambda/\sigma)^2}
\eea
At large distances, taking the small $k$ limit of the expressions
(\ref{eq:Uij}), 
$\ell^{-1}\ll k\ll a^{-1}$,
we obtain
\be
D_2(k)=  
\frac{\alpha}{k^2}
\,,\quad
D_1(k)
= \beta
\ee
where $\alpha=d^2/(8\sigma a^2)$, $\beta=\pi d^2/a$,
and $\ell =(8\pi\sigma a/\lambda)^{1/2}$ is the characteristic length 
of plasmon decay due to vertical leakage across the quantum wells.
The term $D_2$, which has a long-range form
$-\frac{\alpha}{2\pi}\ln (|\vec r-\vec r'|/\ell)$ in position representation,
describes the plasmon wind effect. 
The term $D_1$ describes 
the static inter-exciton interaction screened by charges in the doped regions.
Altogether, we obtain
an expression
\be\label{eq:Ueff(r,t)}
U_{\rm eff}(\vec r)=\frac{\pi d^2}{a} n_{\rm x}(\vec r)
-\frac{\alpha\gamma}{2\pi} \int \ln \lp \frac{|\vec r-\vec r'|}{\ell}\rp
n_{\rm x}(\vec r') d^2r'
\ee
valid at the length scales from $a$ to $\ell$.
In passing from Eq.(\ref{eq:Ueff(k,w)}) to Eq.(\ref{eq:Ueff(r,t)}),
we replaced $\dot  n_{\rm x}$ by $-\gamma n_{\rm x}$, 
with $\gamma$ the recombination rate.
(We note that a similar plasmon shakeup accompanies exciton formation 
by electron and hole binding, leading to qualitatively similar 
inter-exciton attraction.)

To see how the long-range attractive part of $U_{\rm eff}$ generates 
instability, we consider exciton dynamics
\bea\label{eq:dynamics}
&& \p_t n_{\rm x} +\nabla \vec j_{\rm x} = -\gamma n_{\rm x} + J
\\
&& \vec j_{\rm x} = - D\nabla n_{\rm x} - \mu_{\rm x} n_{\rm x}\nabla U_{\rm eff}
\eea
where $\mu_{\rm x}$ is exciton mobility. 
In a simple, although somewhat artificial situation 
when the excitons are generated uniformly over the area,
the homogeneous state is characterized by constant density 
$\bar n_{\rm x}=J/\gamma$. 
Linearizing Eq.(\ref{eq:dynamics}) about this state, with 
$\delta  n_{\rm x}\propto e^{\omega t}e^{i\vec k\vec r}$,
and setting $\omega \le 0$, we obtain the instability criterion:
\be
(D+\mu_{\rm x} \bar n_{\rm x} \beta) k^2+\gamma 
\le \frac{\alpha\gamma \mu_{\rm x} \bar n_{\rm x} k^2}{k^2+\ell^{-2}}
\ee
This condition is simplest to analyze in the absence of leakage, $\ell=\infty$.
Then for instability one must have
\be
\alpha \mu_{\rm x} \bar n_{\rm x}\ge 1
\ee
Replacing conductivity in $\alpha=d^2/(8\sigma a^2)$ by $e^2\mu_d n_d$, where
$\mu_d$ is the mobility and $n_d$ is the carrier density in the doped region,
we obtain the criterion
\be\label{eq:instab_criterion}
d^2 \mu_{\rm x} \bar n_{\rm x}\ge 8 e^2 a^2 \mu_d n_d
\ee
In a coupled quantum well structure with well separation $w$
the dipole moment is given by $d^2=e^2 w^2/\epsilon_{\rm GaAs}$
where $\epsilon_{\rm GaAs}$ is the dielectric constant.

The criterion (\ref{eq:instab_criterion}) shows that the instability 
can be reached by increasing exciton number $\bar n_{\rm x}$
over a threshold value 
$n_{\rm x,c}=8\epsilon_{\rm GaAs} (a/w)^2(\mu_d/\mu_{\rm x}) n_d$.
In the geometry~\cite{Butov02,Butov03}, the ratio $a/w\simeq 10$.
With the factor $8\epsilon_{\rm GaAs}\simeq 10^2$, 
even for the in-plane mobility $\mu_{\rm x}$ orders of magnitude 
higher than the mobility in the doped region, the 
critical density $n_{\rm x,c}$ is quite high. 

While this difficulty might be overcome by employing a stronger
long-range force, e.g. due to phonons, a more serious 
issue with 
this mechanism arises from the consideration of temperature dependence.
In the experiment~\cite{Butov02,Butov03},
the instability occurs at temperature below few Kelvin.
To explain this, the above model
would have to account for strong temperature dependence 
of transport properties such as that taking place in the low conductivity
regime of electron hopping transport.
In constrast, the kinetic effects discussed above lead in a natural way to
a low temperature phase transition which occurs as a precursor of 
exciton degeneracy
without any specific assumptions about the nature of transport.

\section{Conclusion}

In summary, we considered a scenario when 
the realization of quantum degeneracy in a 
cold electron/hole-exciton system is signalled by the development of a 
spatial density modulation. Although our discussion is motivated by the 
photoexcited CQW system in which electrons and holes are spatially separated,
the mechanism is quite general applying also to 
geometries where the electron and hole sources $J_{\rm e,h}$ have a
spatially independent profile.
The instability mechanism appears to depend 
sensitively on there being a \emph{strongly} nonlinear dependence of the 
electron-hole binding rate on the exciton density pointing to the 
importance of stimulated scattering. Although on a qualitative level 
this model is consistent with experimental observations 
of a low temperature transition, 
before considering it as the only explanation more experiments are needed
in order to provide a quantitative assessment of the instability mechanism. 

{\sc Acknowledgement:} We are indebted to Peter Littlewood, Alex Ivanov
and Daniel Chemla
for valuable discussions.
The work at MIT was supported in part by the MRSEC Program of the National
Science Foundation under award number DMR 02-13282.



\end{document}